\documentclass[a4paper, 12pt]{article}
\usepackage{epsfig}
\usepackage{graphicx}
\usepackage{amsmath}

\author{Juli\'an Candia$^{a}$ and Ezequiel V. Albano$^{b}$\\{}\\
$^a${\small\it Center for Complex Network Research and Department of Physics,}\\  
{\small\it Northeastern University, Boston, MA 02115, USA}\\  
{\small Email address: jcandia@nd.edu}\\
$^b${\small\it INIFTA, CCT La Plata,    
 Universidad Nacional de La Plata,}\\ {\small\it La Plata, Argentina}\\
{\small Email address: ealbano@inifta.unlp.edu.ar}}

\title{The Magnetic Eden Model}

\begin{document}
\maketitle

\begin{abstract}
In the magnetic Eden model (MEM), particles have a spin and grow in contact with a thermal bath. 
Although Ising-like interactions affect the growth dynamics, deposited spins are frozen and not allowed 
to flip. This review article focuses on recent developments and future prospects, such as spontaneous 
switching phenomena, critical behavior associated with fractal, wetting, and order-disorder phase transitions, 
the equilibrium/nonequilibrium correspondence conjecture, 
as well as dynamical and critical features of the MEM defined on complex network substrates.
\end{abstract}

\vspace{0.4 true cm}

\section{Introduction}
Half a century ago, a seminal article by Murray Eden presented 
a stochastic kinetic model for the growth of bacterial colonies \cite{ede58}. 
Except for a follow-up publication by the same author \cite{ede61}, 
this model went unnoticed for more than two decades, until it was rediscovered as the 
precursor (and probably the simplest realization) of a large class of kinetic growth models 
that later drew huge attention in many areas of scientific research and technology. 

Indeed, the Eden model and other kinetic growth models such as directed percolation, ballistic 
deposition, diffusion limited aggregation, random deposition with and without relaxation, 
cluster-cluster aggregation, etc., have been applied to a wide variety of phenomena: 
crystal and polycrystalline growth, molecular beam epitaxy, gelation, 
fracture propagation in solids, colloids, dielectrics, epidemic spreading, 
bacterial and fungi growth colonies, sedimentation, vapor deposition, 
wound healing and tissue regeneration, wetting and surface diffusion, etc. 
(for reviews, see, e.g., \cite{her86,fam91,bun91,bun95,bar95,mar96,ede97,hin00,odo04}). 
More recently, natural extensions to the Eden and other standard kinetic growth models 
have been proposed and studied: magnetic \cite{aus93} and charged-particle Eden cluster growth 
\cite{iva99}, binary mixture growth with competition \cite{sai95}, 
magnetic diffusion-limited aggregation \cite{van95}, magnetically controlled ballistic 
deposition \cite{tro03}, Cayley and diluted Cayley trees with two-state particles \cite{van96}, etc. 

In this context, this review article will focus on recent developments on the so-called magnetic 
Eden model (MEM), defined as an aggregate of particles with a magnetic moment coupled through Ising-like 
interactions. The system grows in contact with a thermal bath, but deposited spins are frozen and not allowed 
to flip. Hence, although the interaction configuration energy for MEM clusters is chosen to resemble the 
Ising Hamiltonian, the MEM is a model for irreversible growth under far-from-equilibrium conditions.   

In regular lattices, the MEM's growth process leads to Eden-like self-affine growing interfaces and 
fractal cluster structures in the bulk \cite{van94,aus95}, and displays a rich variety of 
nonequilibrium phenomena, such as thermal order-disorder continuous phase transitions \cite{can01a},  
spontaneous magnetization reversals \cite{can01b}, as well as morphological \cite{can02c,can02b}, 
wetting \cite{can00,can02a}, and corner wetting transitions \cite{man05}. While the MEM was originally 
motivated by the study of structural properties of magnetically textured materials \cite{aus93,van94}, 
this model can also provide useful insight into kinetic phenomena of great experimental and theoretical interest, such as
the growth of metallic multilayers \cite{bov98} and thin films interacting with a substrate \cite{kul06}, 
fluid adsorption on wedges \cite{rej99}, filling of templates imprinted with 
nanometer/micrometer-sized features \cite{jos01,dev03}, etc. The MEM was extensively studied in confined 
stripped geometries \cite{can03} that resemble, for instance, experiments on the growth of quasi-one-dimensional Fe strips 
on Cu(111) vicinal surfaces \cite{she97} and Fe on W(110) stepped substrata \cite{pie00}.  

Despite the conceptual simplicity of its definition, this model displays a remarkably rich behavior and has a great potential 
for further applications, which prompted us to consider the MEM as a kind of ``growing Ising model".  
Indeed, this is more than just a loose assertion: 
a quantitative correspondence between the critical behavior of the Ising model in $d$ dimensions 
and the MEM in confined ($d+1$)-dimensional stripped geometries was conjectured, based on measurements of 
order parameter probability distributions (for $d=1,2$) and critical exponents \cite{can01a}. Remarkably, similar correspondences 
between nonequilibrium two-state systems and the Ising model were independently 
found in other contexts \cite{gri85,dro97}, 
suggesting an intriguing linkage between equilibrium and nonequilibrium systems yet to be better exploited and understood. 

Following the increasing contribution of statistical physics to areas of interdisciplinary interest 
(for reviews, see, e.g., \cite{oli99,wei00,sta06,bor07}),
the MEM was recently studied on small-world and scale-free network substrates  
as a sociophysical model for irreversible opinion spreading phenomena \cite{can06,can07a,can07b}. 
Indeed, physical concepts such as temperature and magnetization, 
spin growth and clustering, ferromagnetic-paramagnetic phase transitions, etc, 
can be meaningfully reinterpreted in sociological/sociophysical contexts.   
As expected, the MEM's dynamical and  critical behavior is observed to depend very strongly on the topology of the 
substrate. Interestingly, however, similarities and differences are 
found when comparing the MEM to analogous equilibrium spin models. 

The rest of the paper is laid out as follows. In Sect. 2, we define the model and outline a useful Monte Carlo simulation procedure. 
In Sect. 3, we provide a brief overview of the main findings about the MEM growing on regular lattice and complex network substrates. 
Finally, in Sect. 4 we state the conclusions and outlook.   

\section{The model: definition and simulation method} 

In the original Eden model, which is defined on a 2D square lattice, 
the growth process starts by adding particles to the immediate neighborhood (the perimeter) 
of a seed particle. Subsequently, particles are stuck at random to perimeter sites, leading to 
the formation of compact clusters with a self-affine interface. The model's behavior is robust, since 
the interface growth exponents are invariant under different seed geometries and different rules for the deposition 
process \cite{jul85,fre85,hir86}. In fact, it is well established that the interface of the Eden model is 
described by the Kardar-Parisi-Zhang (KPZ) equation \cite{kar86}, similarly to other models that 
belong to the KPZ universality class \cite{bar95,hin00}, such as, e.g., ballistic deposition.  

Motivated by the observation of structural features in the magnetically textured growth of high critical-current 
density superconductors \cite{han92, clo92}, the magnetic Eden model (MEM) was originally proposed in Ref.~\cite{aus93} as an extension of 
the Eden model, in which an additional degree of freedom represents the spin of the growing particles. 
Starting from a seed, growth takes place by adding, one by one, further spins to the perimeter 
of the growing cluster, taking into account the corresponding interaction energies. By analogy
to the Ising model, the energy $E$ of a configuration of spins is given by
\begin{equation}
E = - \frac{J}{2}\sum_{\langle ij\rangle} S_iS_j -\sum_i H_iS_i,
\label{energy}
\end{equation}
where $S_i= \pm 1$ indicates the orientation of the spin for each occupied site (labeled by the 
subindex $i$), $J$ is the coupling constant between nearest-neighbor (NN) spins, $H_i$ is the magnetic 
field applied on site $i$, and $\langle ij\rangle$ indicates that the summation is taken over all pairs of occupied NN sites. 

Setting the Boltzmann constant equal to unity ($k_B\equiv 1$), 
the probability for a new spin to be added to the (already grown) cluster is
defined as proportional to the Boltzmann factor exp$(-\Delta E /T)$, 
where $\Delta E$ is the resulting total energy change and $T$ is the absolute temperature of the thermal bath\footnote{Energy, 
magnetic field, and temperature are measured here in units of the NN coupling constant, $J$, throughout.}. Notice that, actually, the 
total energy change involves only the deposition site and its occupied nearest neighbors: the change in the 
spin-spin interaction term (i.e., the first term in Eq.(\ref{energy})) just 
reflects the magnetic coupling between the already existing cluster and the newly added spin, while the change in the second 
term of Eq.(\ref{energy}) is simply given by the interaction between the new spin and the local magnetic field.    
At each step, all perimeter sites have to be considered and the probabilities of adding a new (either up 
or down) spin to each site must be evaluated. 
Using the Monte Carlo simulation method, all growth probabilities are first computed and normalized, and then    
the growing site and the orientation of the new spin are both determined by means of a pseudo-random number\footnote{Fortran codes with 
Monte Carlo implementations of the MEM under different setups are freely available from the authors upon request.}.
Let us point out again that, although Eq.(\ref{energy}) resembles the Ising Hamiltonian, the MEM is a nonequilibrium model in 
which new spins are continuously added, while older spins remain frozen and are not allowed to flip. 

\begin{figure}[t]
\centerline{{\epsfxsize=4.3in \epsfysize=2.9in \epsfbox{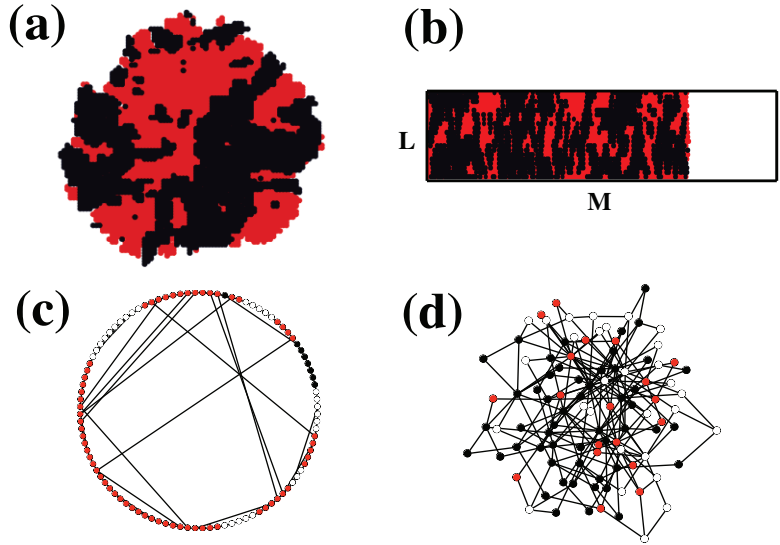}}}
\caption{MEM growth under different setups: (a) planar substrates on the square lattice, (b) ($d+1$)-dimensional 
confined (stripped) geometries, (c) small-world networks, and (d) scale-free networks. Red (black) sites represent up (down) spins, 
while empty sites are shown in white.} 
\label{fig1}
\end{figure}

Figure 1 shows different setups that have already been considered: (a) planar substrates on the square lattice, (b) ($d+1$)-dimensional 
confined (stripped) geometries, (c) small-world networks, and (d) scale-free networks. Up (down) spins are shown in red (black), 
empty sites in white.
The typical Monte Carlo procedure requires performing averages over a large number of statistical ensembles generated 
under the same conditions (i.e., temperature, magnetic fields, NN couplings, etc). For complex network substrates, ensemble averages over 
different (but topologically equivalent) network realizations are necessary as well. 

Note that the substrate geometry naturally suggests the seed to be chosen. For planar substrates (Figure 1(a)), a single spin originates 
unconfined clusters (which, growing on the square lattice, acquire a diamond shape \cite{bat91}). For $L^d\times M$ 
stripped geometries with $L\ll M$ (Figure 1(b)), linear/planar transversal seeds lead to longitudinal growth. For small-world (Figure 1(c)) 
and scale-free (Figure 1(d)) networks, growth from single spin seeds naturally stops once the system becomes completely filled.  
  
Allowing the MEM to grow under different conditions, a remarkably rich variety of growth phenomena is encountered. 
Indeed, different growth modes arise from using confined and unconfined geometries, varying the dimensionality of the substrate, 
considering positive and negative coupling constants, applying surface and bulk magnetic fields (which can also be 
spatially homogeneous, periodic or random), changing the topology of the substrate, etc. Some of these phenomena were 
recently uncovered and will be summarized in the next section. Yet, a horizon of exciting new possibilities for the MEM applied to 
fields as different as materials science, biophysics, and sociophysics can be envisioned. 

\section{Overview of main results}

\subsection{Unconfined geometries: Lacunarity, fractal and\\ magnetic pseudophase transitions} 
  
The 1D MEM growing from a single seed can be solved exactly and leads to a thermal order-disorder pseudophase transition   
taking place at a finite ``critical temperature" $T_c(N)$, which is weakly dependent upon the system size $N$. 
However, since $T_c^{-1}\sim\log(N)$, the system is noncritical in the ($N\to\infty$) thermodynamic limit \cite{aus93}.  
A similar scaling behavior was also reported for the MEM growing on unconfined 2D planar substrates (Figure 1(a)). 
Interestingly, however, these finite-size, order-disorder ``phase transitions" are well correlated to geometrical transitions 
associated with fractal and lacunar structures in the bulk \cite{van94,aus95}. 

\begin{figure}
\centerline{{\epsfxsize=5.4in \epsfysize=2.4in \epsfbox{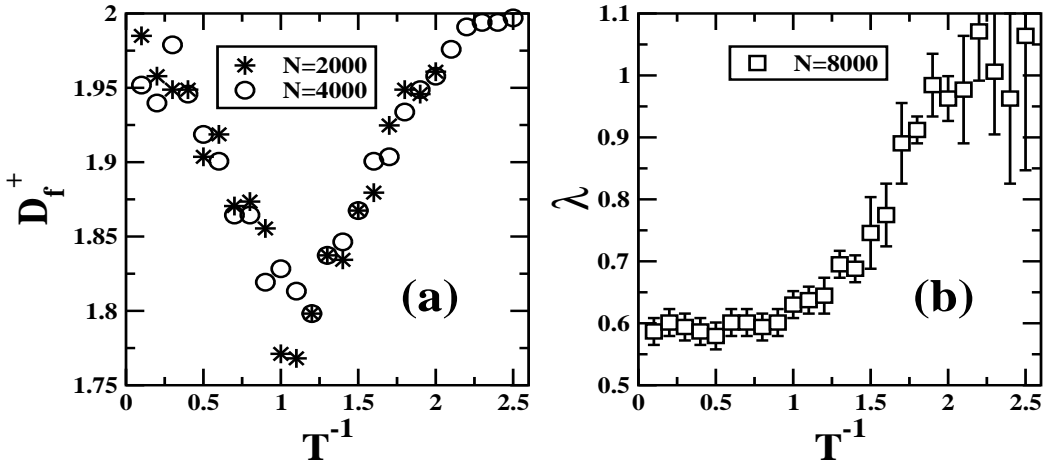}}}
\caption{Geometrical transitions for the MEM growing on 2D planar substrates: 
(a) fractal dimension of the up component versus temperature, and (b) lacunarity exponent versus 
temperature (adapted from Ref.~\cite{van94}).} 
\label{fig2}
\end{figure}

The fractal dimension $D_f^+$ of the up species can be measured by counting the number $n^+$ of up spins in boxes of 
different sizes $\epsilon$ centered on the seed site, i.e.,
\begin{equation}
n^+\sim\epsilon^{D_f^+}\ ,
\end{equation}
a method originally used to obtain the fractal dimension of smoke aggregates \cite{for79}. 
Figure 2(a) shows a clear departure of $D_f^+$ from the Euclidean space dimension of the substrate, $D=2$, 
with a sharp minimum equal to $D_f^+=1.79\pm 0.03$ at the ``critical" temperature $T_c^{-1}=1.2\pm 0.1$ (for $N=4000$) \cite{van94}. 

The lacunarity density, $l_d$, is defined as the total number of empty sites enclosed in the bulk by occupied neighbors, 
normalized by the total number of deposited spins $N$. For the Eden model, the lacunarity density obeys the scaling law
\begin{equation}
l_d\sim N^{\lambda-1}\ ,
\end{equation}
with $\lambda=0.56\pm 0.01$. For the MEM, however, the lacunarity exponent depends on the temperature and undergoes a 
transition towards unity at $T\approx T_c$ \cite{van94}. Figure 2(b) shows the thermal dependence of the lacunarity 
exponent for a system of size $N=8000$.     

\subsection{Confined stripped geometries: spontaneous switching, 
phase transitions and the equilibrium / nonequilibrium correspondence conjecture}
At low temperatures, magnetic Eden films grown on a stripped geometry of finite linear dimension $L$ (Figure 1(b)) display 
{\it spontaneous magnetization reversals}: a sequence of well-ordered magnetic domains 
separated by abrupt collective spin reversals of characteristic length $l_R\sim L$ \footnote{
Notice that, since we consider $L\times M$ strips with $L\ll M$, the longitudinal direction is effectively 
infinite and $L$ is the only relevant linear scale in the substrate.}\cite{can01b} (see Figure 3). 
This phenomenon is due to thermal fluctuations on the finite-size thin films. Indeed, although at low temperatures 
the bulk grows in an ordered state, sizable fluctuations may occur, eventually driving the finite-size system across a drastic 
magnetization reversal. 

The probability of occurrence of spontaneous magnetization reversals vanishes as the system's linear size tends to the 
($L\to\infty$) thermodynamic limit. However, this mechanism is relevant for many real mesoscopic systems, since clearly such 
spontaneous (and, thus, uncontrollable) reversals must be avoided in the preparation of
high-quality magnetic thin films and nanowires. These shortcomings may disappear if the film strongly interacts with the substrate
where the actual growing process takes place, as modeled, for instance, by surface and/or bulk 
fields that could account for the interaction with the substrate or with externally applied magnetic fields.    
 
\begin{figure}
\centerline{{\epsfxsize=3.6in \epsfysize=1.8in \epsfbox{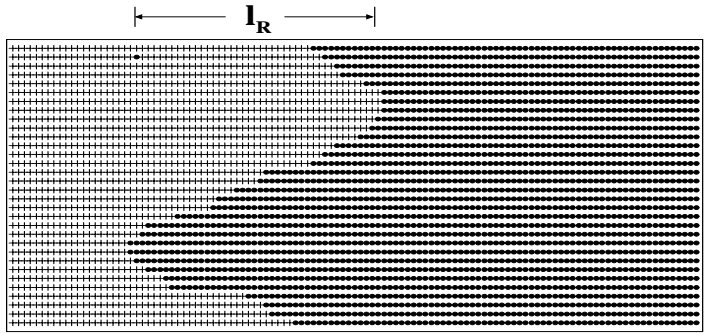}}}
\caption{Spontaneous magnetization reversal in a $(1+1)-$dimensional magnetic thin film. 
The snapshot corresponds to the bulk of the sample and the growing interface is not shown.} 
\label{fig3}
\end{figure}

The degree of order in a magnetic system can be naturally characterized by the ensemble-averaged magnetization per site. 
In the case of ($d+1$)-dimensional stripped geometries, the magnetization is averaged over the transversal direction 
(e.g., transversal columns of $L$ spins for $d=1$, and transversal planes of $L\times L$ spins for $d=2$), i.e.,
\begin{equation}
m=\langle {\frac{1}{L^d}}\sum S_i\rangle \ .
\end{equation}

The thermal dependence of the order parameter probability 
distributions, $P_L(m)$, contains information about all momenta of the order parameter $m$, including
universal ratios such as the Binder cumulant \cite{bin81}.
Figure 4 shows plots of $P_L(m)$ versus $m$ for different values of temperature, substrate dimension and linear size: 
(a) $d=1, L=128$; and (b) $d=2, L=16$. 
In both cases, the high-$T$ distributions are peaked at $m=0$ and can be well approximated by Gaussian distributions. However, as 
temperature decreases, their behavior is fundamentally different. In the $d=1$ case, 
the distribution broadens and develops two peaks at $m=\pm 1$, which  
become dominant while the distribution turns distinctly non-Gaussian, with an absolute minimum at $m=0$. In sharp contrast, 
by lowering the temperature in the $d=2$ case the distribution develops two symmetrical maxima located at $m=\pm M_{sp}$ $(0<M_{sp}<1)$,
which become steeper and approach $m=\pm 1$ as $T$ decreases further. 

This behavior bears close resemblance to the order parameter probability
distributions of the $d-$dimensional Ising model \cite{lan80,bin96}, suggesting a nontrivial connection 
between the critical behavior of the ($d+1$)-dimensional MEM and the Ising model in $d$ dimensions (see \cite{can01a} 
for a full discussion). For $d=1$, the ordered phase is trivially found only at $T=0$, while for $d=2$, 
MEM films exhibit a continuous order-disorder phase transition at the critical temperature $T_c = 0.69 \pm 0.01 $. 
This connection can indeed be further confirmed by measurements of critical exponents in the $d=2$ case. 
Using standard finite-size scaling analysis \cite{bar83,pri90}, the MEM's critical exponents are found: 
$\nu = 1.04 \pm 0.16 $, $\gamma = 2.10 \pm 0.36$, and $\beta = 0.16 \pm 0.05$,  
which agree well within error bars with the exact critical exponents for the Ising model in 2 dimensions: 
$\nu=1$, $\gamma=7/4$ , and $\beta=1/8$. 

\begin{figure}
\centerline{{\epsfxsize=5.5in \epsfysize=2.3in \epsfbox{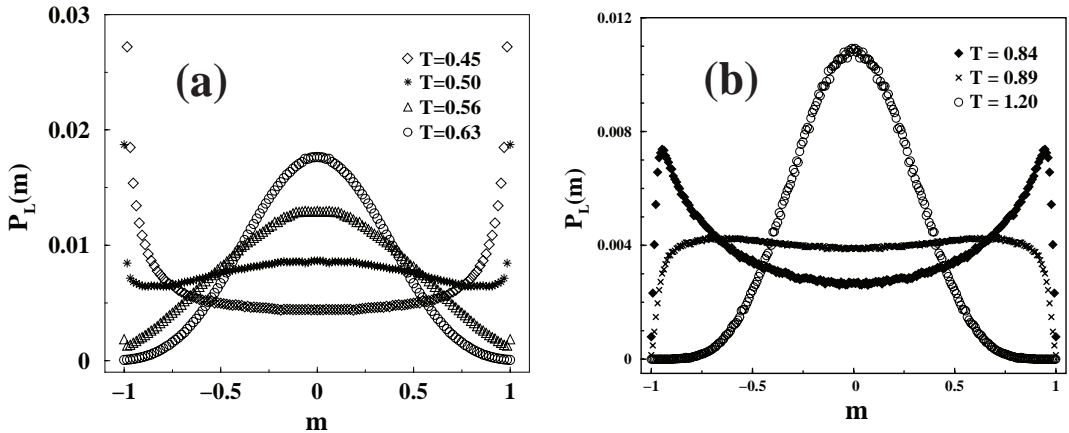}}}
\caption{Order parameter probability distributions for different temperatures and substrates: 
(a) $d=1, L=128$; (b) $d=2, L=16$. In (a), 
the sharp peaks at $m =\pm 1$ for $T=0.45$ have been truncated.}
\label{fig4}
\end{figure}

It is interesting to point out that 
similar connections between nonequilibrium systems and the Ising model were independently 
found in other contexts. For $d-$dimensional probabilistic cellular automata with 
two states and up-down symmetry, it was shown that if the system undergoes a symmetry breaking, 
its critical behavior is identical to the corresponding 
Ising model in equilibrium \cite{gri85}. Furthermore, studies on thin film ($d+1$)-dimensional binary alloys 
with surface equilibration (but such that atoms in the bulk are frozen) show that the behavior parallel to the growth 
direction reflects the dynamics of an Ising system in $d$ dimensions \cite{dro97}. 
These findings suggest an intriguing linkage between equilibrium and nonequilibrium systems that should 
be further exploited and understood, and the MEM offers an open avenue to gain deeper insight into this issue. 

Certainly, the validity of this conjecture could be further confirmed (or, for that sake, otherwise ruled out) by performing 
more accurate and complete numerical studies that may narrow down error bars and explore other critical exponents. 
In the same vein, the exploration of higher dimensional systems (i.e., $d\geq 3$) would contribute to this goal. 
However, MEM's growth rules require updating the growth probabilities at each time step and lead to very slow 
algorithms compared to analogous equilibrium spin models. An alternative path is the search for analytic solutions, 
so far only known for the $d=1$ semiopen chain \cite{aus93}. 

\begin{figure}
\centerline{{\epsfxsize=5.5in \epsfysize=2.5in \epsfbox{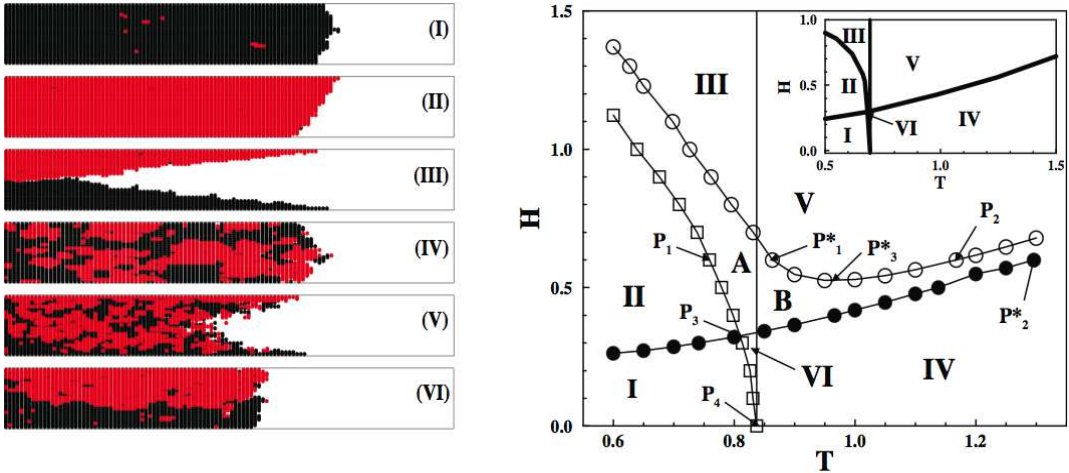}}}
\caption{Bulk and interfacial phase transitions in $L^2\times M$ thin films. 
The left panel shows longitudinal slice snapshots for $L=32$ and different temperatures and surface magnetic 
fields: (I) $H=0.05$, $T=0.6$; (II) $H=0.5$, $T=0.55$; (III) $H=1.4$, $T=0.6$; (IV) $H=0.1$, $T=1.0$; (V) $H=1.6$, $T=1.4$;
and (VI) $H = 0.20$, $T = 0.82$. The right panel shows the $H-T$ phase diagram for $L=12$, as well as the extrapolation to
the thermodynamic limit (inset).}
\label{fig5}
\end{figure}

Interfacial phase transitions and critical phenomena in confined samples exhibit a quite
distinct physical behavior compared to that occurring in the bulk due to the finite separation between the walls
and the specific wall-particle interaction \cite{fis81}. When competing short-range magnetic fields are applied to 
the upper and lower surfaces along the longitudinal direction, MEM films exhibit different kinds of growth mode and 
lead to a rich phase diagram on the $H$ versus $T$ plane (see \cite{can02a} for a full discussion). 
Indeed, the interplay between confinement
and growth mode leads to a localization-delocalization transition in the interface that runs along the walls and a 
change of the curvature of the growing interface running perpendicularly to the walls. Extrapolation
of this scenario to the thermodynamic limit leads to a multicritical wetting 
transition under far-from equilibrium conditions. 
It is worth mentioning that the study of out of equilibrium wetting phenomena
has also attracted considerable attention in a context related to the evolution of
growing interfaces \cite{hin97,tu97,kis05}.

The occurrence of different growth modes is evident in the snapshot configurations on the left panel of Figure 5, where 
longitudinal slices of $L^2\times M$ films ($M\gg L$) are shown for different temperatures and surface magnetic fields.  
As above discussed, thermally driven transitions in the bulk separate the low-$T$ ordered regime (phases I, II, III, and VI) 
from the high-$T$ disordered one (phases IV and V). Moreover, a wetting localization-delocalization phase transition 
(associated with the longitudinal interface between up and down domains) takes place within the ordered-bulk phase, 
separating the nonwet region (phases I and II) from the wet region (phases III and VI). Finally, due to the so-called 
missing neighbor effect and varying intensity of wall-particle interactions, different contact angles and curvatures in 
the {\it growing} interface 
(i.e., the transverse interface between occupied and empty lattice sites) are also observed. These wetting-like, 
morphological transitions separate distinct growth regimes, characterized by either convex (phases I and IV), nondefined 
(phases II and VI), or concave (phases III and V) growing interfaces. 

Standard procedures allow quantifying these phenomena, leading to the rich phase diagrams shown on the right panel 
of Figure 5. For finite-size systems, two additional regions (labeled $A$ and $B$) are observed, although they just 
represent intermediate states that cannot be associated with distinct physical processes. By extrapolating to the thermodynamic 
limit, regions $A$ and $B$ shrink and vanish, leading to the phase diagram shown in the inset.  
In different geometrical settings, such as, e.g., growth on wedges \cite{man05}, other interesting interfacial phase transition phenomena 
are observed (not discussed here for the sake of space).        

\subsection{Complex networks: dynamics and criticality of\\ opinion spreading phenomena}

As part of the increasing application of statistical physics to interdisciplinary fields of science, 
a variety of Ising-like spin models have been recently studied in the context of sociophysical 
phenomena (see, e.g., \cite{bar00,git00,bor01,her02,ale02,sve02,med03,her04}). 
Under this interpretation, spin states denote different opinions or preferences, while the  
coupling constant describes the convincing power between interacting individuals, 
which is in competition with the ``free will'' represented as thermal noise \cite{ale02}.
In particular, much attention has been devoted to  
spin models defined on complex networks, since the underlying complex network topology 
of the substrate reflects some key aspects of social structures, such as the small-world effect and the 
high connectivity of local neighborhoods \cite{wat99,alb02,dor03,new06}. 
  
In this context, the MEM has been recently studied on small-world and scale-free network substrates 
as a sociophysical model for irreversible opinion spreading phenomena \cite{can06,can07a,can07b}. 
According to the MEM's rules, the opinion or decision of an individual is affected by those of his or her acquaintances, 
but opinion changes (analogous to spin flips in an Ising-like model) do not occur.
Hence, as opposed to equilibrium spin models, the MEM 
could be applied to sociological scenarios in which individuals are subject 
to highly polarized, short term, binary choice situations. Given these conditions opinions are not expected to fluctuate 
and ``thermalize". One example is a binary voting scenario, such as a ballotage or referendum, where model predictions 
could be empirically tested using time-resolved data from polls and surveys.

The MEM was studied on two paradigmatic types of complex network: (a) the nearest-neighbor, adding-type 
small-world network (SWN) \cite{bol88,new99b}, which 
is built by randomly adding new links onto an underlying ordered lattice
(for each bond in the original lattice, a shortcut is added with probability $p$); (b) Barab\'asi-Albert (BA) and 
uncorrelated scale-free (SF) networks, which are characterized by power-law degree distributions, i.e., $P(k)\sim k^{-\gamma}$.
BA networks are built by following the preferential attachment growth mechanism: a new node with $m$ edges   
is added at every time step and is connected to $m$ different nodes already present 
in the system. The probability for an already 
existing node to acquire a new link is proportional to that node's degree \cite{alb02,bar99}. Instead, 
uncorrelated SF networks arise from pairwise connecting link ends, which are 
chosen according to a power-law distribution \cite{alb02,dor03}. 
The degree exponent for BA networks is $\gamma_{BA}=3$, while in the 
case of uncorrelated SF networks, $\gamma$ is a free parameter.   

\begin{figure}
\centerline{{\epsfxsize=5.7in \epsfysize=2.6in \epsfbox{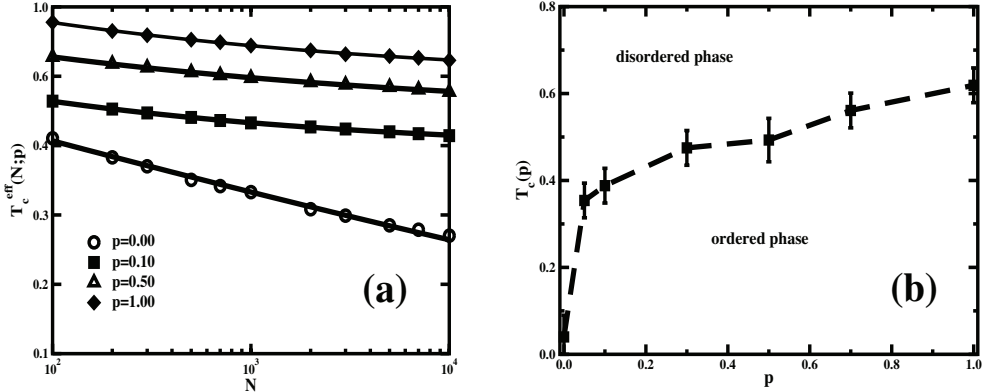}}}
\caption{(a) Effective transition temperatures for the MEM growing on 1D SWNs of size $10^2\leq N\leq 10^4$ and different 
values of $p$ (symbols), and fits to the data (solid lines) using the finite-size scaling relation, 
Eq.(\ref{scaling1}). (b) Phase diagram $T_c(p)$ versus $p$ corresponding to $N\to\infty$.}
\label{fig6}
\end{figure}

The symbols in Figure 6(a) show effective transition temperatures, $T_c^{eff}(N;p)$, 
which separate the ordered regime from the disordered one for the MEM growing on 1D SWNs of different size and 
shortcut-adding probabilities. Fits to the data are also shown (solid lines), as  
obtained from the finite-size scaling relation
\begin{equation}
|T_c(p)-T_c^{eff}(N;p)|\propto N^{-1/\nu}, 
\label{scaling1}
\end{equation}
where $T_c(p)$ is the true $p-$dependent critical temperature (corresponding to $N\to\infty$) and 
$\nu$ is the exponent that characterizes the divergence of the correlation length 
at criticality. In Figure 6(b), the phase diagram $T_c(p)$ versus $p$ shows the critical behavior of the system in 
the thermodynamic limit.
For $p>0$, the MEM undergoes critical order-disorder phase transitions at finite critical temperatures: 
the small-world network geometry triggers criticality. Naturally, 
the weaker the global order imposed by long-range shortcuts, the 
lower the shortcut fraction, and hence $T_c(p)$ decreases monotonically with $p$. Moreover, the critical temperature 
is observed to vanish for $p=0$, which is the expected regular lattice limit behavior. 

The same fits of Eq.(\ref{scaling1}) to the numerical data also determine the critical exponent $\nu$. 
The obtained ($p-$independent) value is $\nu=3.6\pm 0.4$. 
As in the case of equilibrium spin systems defined on small-world networks (see, e.g., \cite{her02,med03}), 
the universality class 
of this nonequilibrium system is not observed to depend on the shortcut density, provided that $p>0$.  
An additional characterization of the critical behavior of the MEM defined on 1D SWNs can be obtained 
by calculating the critical exponent $\gamma$, which describes the divergence of the susceptibility 
at the critical point. The exponent ratio $\gamma/\nu$ can 
be related to the peak of the susceptibility measured in finite samples of size $N$ by \cite{pri90}
\begin{equation}
\chi_{max}\propto N^{\gamma/\nu}. 
\label{scaling2}
\end{equation}
It turns out that $\gamma/\nu=0.92\pm 0.04$, and hence $\gamma=3.3\pm 0.4$. 

\begin{figure}
\centerline{{\epsfxsize=5.7in \epsfysize=2.6in \epsfbox{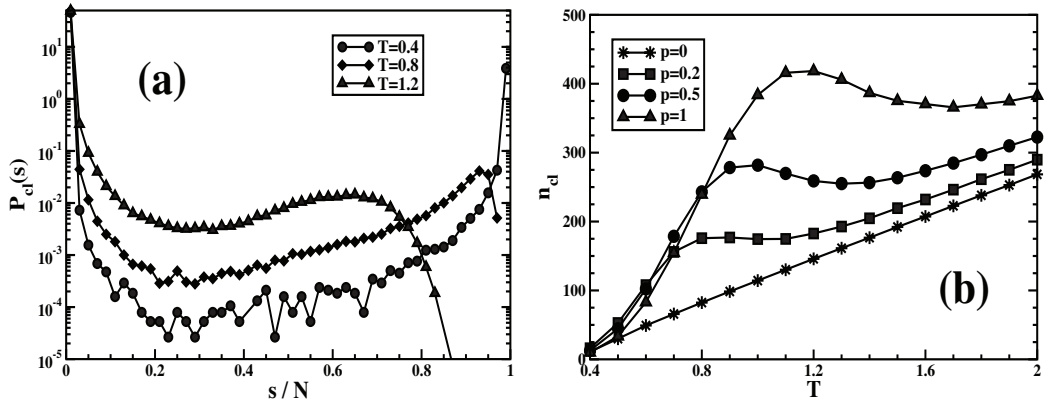}}}
\caption{Domain structure for the MEM growing on 2D SWNs of size $N=10^4$ and different values of 
shortcut-adding probability and temperature: (a) cluster size probability distribution (for $p=0.5$); 
(b) mean number of different clusters.} 
\label{fig7}
\end{figure}

The onset of criticality induced by the presence of long-range connections is also observed when the MEM grows 
on 2D SWNs \cite{can07b}. The underlying 2D substrate, however, allows observing other phenomena of interest, such 
as the collision of different shortcut-induced opinion fronts, domain growth and cluster formation. 
The formation of rich domain structures, which depends on both the temperature and the shortcut distribution, can 
be characterized quantitatively by measurements such as the cluster size probability distribution, the size of the largest cluster, 
and the number of different clusters. Figure 7 shows the cluster size probability distribution and the mean number 
of different clusters for a system of size $N=10^4$ and different values of 
shortcut-adding probability and temperature.      

The probability of occurrence of a cluster of size $s$, $P_{cl}(s)$, is shown in Figure 7(a) for $p=0.5$ and different 
temperatures, as indicated. Due to thermal 
fluctuations, all distributions show an absolute maximum at small cluster sizes ($s/N\ll 1$). However, the probability 
distribution for large clusters reveals intrinsic differences in the growth mode, which depend on the temperature. 
Apart from statistical fluctuations, the distributions of clusters of size $s/N\geq 0.2$ grow monotonically with $s$. 
However, for $T=0.4$, the monotonic trend continues up to $s/N=1$, where the distribution has a local maximum, 
while for larger temperatures, the distributions show abrupt cutoffs at $s/N<1$, the more evident, the larger the temperature. 
Figure 7(b) shows the thermal dependence of the mean number of different clusters, $n_{cl}$. 
The $p>0$ plots show broad local maxima taking place at intermediate temperatures. As with response functions such as 
the susceptibility and heat capacity, the observed maxima are indicative of   
peak fluctuations in the cluster structure due to thermally driven, bulk order-disorder phase transitions. Indeed, following 
a procedure analogous to that described above for 1D SWNs, one can characterize the critical behavior of the system in 
the thermodynamic limit (see \cite{can07b} for more details). 

\begin{figure}
\centerline{{\epsfxsize=5.7in \epsfysize=2.6in \epsfbox{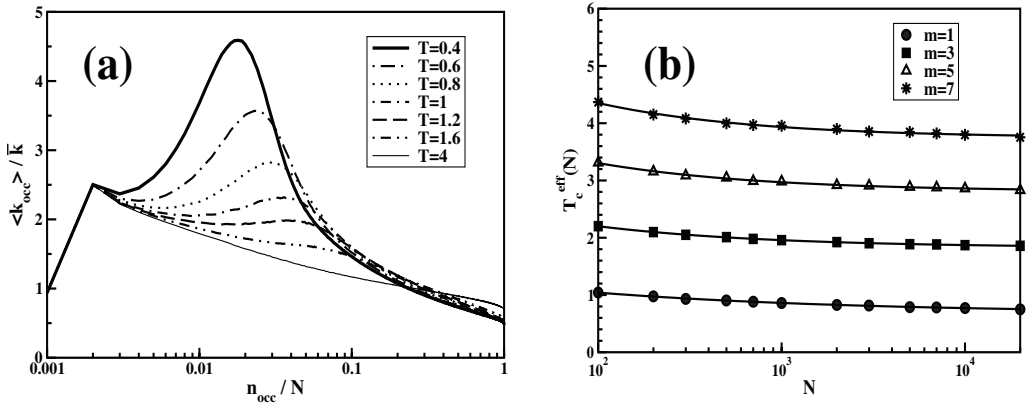}}}
\caption{(a) Average degree of occupied nodes for the MEM growing on BA networks 
of size $N=10^3$, $m=3$, and different temperatures, as a function of the 
fraction of deposited particles, $n_{occ}/N$. (b) Effective transition temperatures corresponding to BA networks of different 
sizes and values of the parameter $m$ (symbols), and finite-size scaling fits (solid lines).}  
\label{fig8}
\end{figure}

Contrary to the case of SWNs, SF networks provide an interesting setting in which the social roles 
differ significantly from one agent to another. In particular, hubs in the SF network 
represent highly influential individuals, whose opinions can potentially affect the decisions of many other 
individuals in the society.  
Let us first discuss the dynamical behavior of MEM clusters growing on finite BA scale-free networks.   

The process of spin deposition (or, equivalently, opinion spreading) can be characterized by computing the degree of the 
newly occupied node, $k_{occ}$, each time a new particle is added to the system.
Figure 8(a) shows the average degree of occupied nodes relative to the network's mean degree, $\langle k_{occ}\rangle / \bar{k}$, 
for BA networks of size $N=10^3$, $m=3$, and different temperatures, as a function of the 
fraction of occupied nodes, $n_{occ}/N$. Since at low temperatures the system tends to  
develop highly ordered spin domains, highly connected nodes have larger probabilities to be occupied at early times
during the growth process. The leading role of hubs is clearly observed in the low-temperature 
plots of Figure 8(a), where $\langle k_{occ}\rangle / \bar{k}\gg 1$ at early stages of the growth process, i.e.,  
when the number of occupied nodes is of the order of a few percent of the total system size. At higher temperatures, however, 
the increased thermal noise tends to wash out the phenomenon of {\it preferential spin deposition}. The growth at later 
times mainly proceeds by the occupation of less-than-average connected nodes, which leads to roughly $T$-independent 
values of $\langle k_{occ}\rangle$. 

\begin{figure}[t]
\centerline{{\epsfxsize=5.7in \epsfysize=2.6in \epsfbox{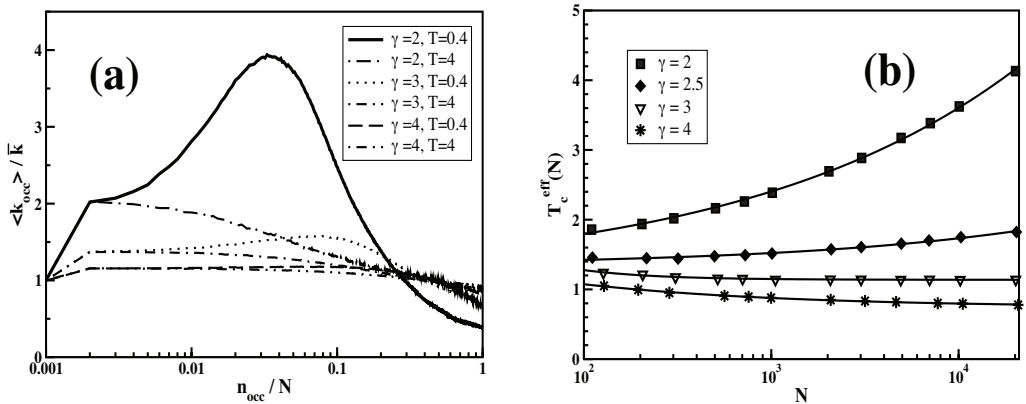}}}
\caption{(a) Average degree of occupied nodes for uncorrelated SF networks 
of size $N\approx 10^3$, minimum degree $k_0=3$, and different values of temperature and degree exponent.
(b) Effective transition temperatures for uncorrelated SF networks (with $k_0=3$) as a function of size and degree exponent
(symbols), along with finite-size scaling fits (solid lines).}
\label{fig9}
\end{figure}

The critical behavior of the system, as follows from the standard finite-size scaling procedures discussed above, is 
depicted in Figure 8(b). The effective pseudocritical temperatures decrease monotonically with the network size and lead to 
finite extrapolations in the thermodynamic limit.  
The monotonic increase of $T_c(m)$ can be well 
approximated by the linear relation $T_c(m)= 0.522(3)\times m +0.21(1)$, while the exponent
$\nu$ decreases monotonically and tends to $\nu\approx 2$ for $m\geq 3$.    

Finally, let us consider the irreversible growth of MEM clusters on uncorrelated SF networks, which 
allow a free choice of the degree exponent $\gamma$. 
Figure 9(a) shows $\langle k_{occ}\rangle / \bar{k}$ versus $n_{occ}/N$  
for uncorrelated SF networks generated with different values of $\gamma$ and for different temperatures, as indicated.   
The phenomenon of preferential spin deposition, 
due to the dominant role played by hubs during the early stages of the growth process, is only significant 
at low temperatures. Moreover, this phenomenon is observed to be relevant only for low degree exponents $\gamma<3$. 
Since the preferential spin deposition is a feature associated with the formation of large ordered clusters during 
the growth process, these results agree well, at a qualitative level, with the analogous behavior reported for the 
Ising model on uncorrelated scale-free networks \cite{her04}, in which the disorder was observed to grow 
monotonically with the exponent $\gamma$.  

Figure 9(b) shows the effective transition temperatures for uncorrelated SF networks of 
minimum degree $k_0=3$, for different values of network size and the degree exponent. 
The critical behavior turns out to be crucially dependent on the steepness of the degree distribution:  
while for $\gamma\geq 3$, the trend is similar to the behavior observed for BA networks, the plots corresponding 
to smaller values of $\gamma$ are observed to diverge, hence implying the absence of paramagnetic-ferromagnetic phase transitions in 
the thermodynamic limit. 

Using finite-size scaling procedures, one finds $T_c=1.14(2)$ and $\nu=0.9(1)$ (for $\gamma=3$), and 
$T_c=0.74(1)$ and $\nu=2.6(2)$ (for $\gamma=4$). 
Comparing these observations with the results reported for the Ising model on uncorrelated scale-free 
networks \cite{her04}, we find a qualitative agreement for both $\gamma<3$ (i.e., the absence of 
a paramagnetic phase in the thermodynamic limit) and $\gamma>3$ (i.e., the existence of a finite critical temperature 
delimiting the paramagnetic-ferromagnetic phase transition). 
Instead, the convergence of the critical temperature for the $\gamma=3$ case is clearly in contrast with the 
logarithmic divergence observed in the Ising model. 

\section{Conclusions and Outlook}

The magnetic Eden model (MEM) is a kinetic growth model in which particles have a spin and grow in contact with a thermal bath. 
Although Ising-like interactions affect the growth dynamics, deposited spins are frozen and not allowed 
to flip. This model describes nonequilibrium binary mixture growth phenomena that have a wide range 
of potential applications in contexts as different as materials science, sociophysics, and biophysics. 
 
In regular lattices, the MEM's growth process leads to Eden-like self-affine growing interfaces and 
fractal cluster structures in the bulk, and displays a rich variety of 
nonequilibrium phenomena, such as thermal order-disorder continuous phase transitions,  
spontaneous magnetization reversals, as well as morphological, wetting, and corner wetting transitions.

In a sense, the MEM can be regarded as a ``growing Ising model", and indeed 
a quantitative correspondence between the critical behavior of the Ising model in $d$ dimensions 
and the MEM in confined ($d+1$)-dimensional stripped geometries was conjectured, based on measurements of 
order parameter probability distributions and critical exponents. Certainly, further work in this direction is 
needed to gain further understanding of this correspondence, which  
suggests an intriguing linkage between equilibrium and nonequilibrium systems. Remarkably, similar connections 
between nonequilibrium two-state systems and the Ising model were independently 
found in other contexts.  

Very recently, the MEM has also been investigated on complex network substrates such as small-world and scale-free 
networks. Interpreted as a sociophysical model for irreversible opinion spreading phenomena, the MEM offers a 
complementary view of sociological processes studied by means of other spin models, and it could be applied 
to scenarios in which individuals are subject to highly polarized, short-term, binary choice situations, such as 
a ballotage or referendum. Other applications in this area may also include marketing campaigns, in which the influence of advertising 
could be modeled by externally applied fields.  

This review summarizes the main results obtained so far on the MEM, with a bias towards the most recent work and applications, 
as well as discussions on open questions and outlook. 
Hopefully, growing research efforts on the MEM will follow, stimulating and contributing to further developments in the fields of 
nonequilibrium statistical physics, complex networks, and interdisciplinary science. 

\section*{Acknowledgments}
This work is partially supported by CONICET, ANPCyT, and UNLP (Argentina).
J. C. is supported by the James S. McDonnell Foundation and the National 
Science Foundation ITR DMR-0426737 and CNS-0540348 within the DDDAS program.

\end{document}